\documentclass[12pt]{amsart}
\usepackage{geometry}                % See geometry.pdf to learn the layout options. There are lots.
\usepackage{graphicx}
\usepackage{amssymb}
\usepackage{epstopdf}
\title{Why Colloidal Systems can be described by Statistical Mechanics:\\Some not very original comments on the Gibbs paradox}
\author{Daan Frenkel $^{\ast}$
\\\vspace{6pt}  {\em{Department of Chemistry\\ University of Cambridge\\ Cambridge CB2 1EW, UK}}}
\thanks{$^\ast$Email: df246@cam.ac.uk}
\begin{document}
\maketitle
%\section{}
%\subsection{}

\begin{abstract}
Colloidal particles are distinguishable. Moreover, their thermodynamic properties are extensive.
Statistical Mechanics predicts such behaviour if one accepts that the configurational integral of a system of $N$ colloids must be divided by $N!$. 
In many textbooks it is argued that the factor $N!$ corrects for the fact that identical particles (in the quantum mechanical sense) are indistinguishable. Clearly, this argument does not apply to colloids. This articles explains why, nevertheless, all is well. The point has been made before, but has not yet sunk in. I also discuss the effect of polydispersity.
\end{abstract}

\section{Background} 
There should be no need to write this article about the Gibbs Paradox, but I am afraid that there is.
The Gibbs paradox is based on the observation that when two systems of identical particles in the same thermodynamic state are brought into contact, the entropy of the combined system does not change. Yet the entropy does increase when we allow mixing of two systems of `almost' identical particles that we have somehow managed to separate. 

In Gibbs's classical Statistical Mechanics, the entropy of a system with a fixed constant number of particles (N), volume (V) and energy (E) is related to the logarithm of $\Omega(N,V,E)$, the volume in phase space accessible to this system. In Boltzmann's (posthumous) notation:
\begin{equation}\label{eq:SkLnOmega}
S=k_B\ln\Omega(N,V,E)
\end{equation}
Gibbs (and, before him,  Planck) realised that this expression only defines the entropy up to a constant that does not depend on $V$ or $E$, but can depend on $N$.
In fact, Boltzmann never wrote down equation~\ref{eq:SkLnOmega}.  Planck did~\cite{Planck}, acknowledging Boltzmann's influence. 
However, Planck wrote $S_N=k \log W + \mbox{const.}$  Somehow, the constant got lost in translation. Probably because, by the time the famous text was written on Boltzmann's grave~\footnote{Apparently the text was proposed by Max Planck, around 1930.}, quantum mechanics, and the `quantum' interpretation of Nernst 3$^{rd}$ Law of Thermodynamics~\footnote{Based on very limited statistics (2), it seems that 3$^{rd}$ Laws do not have the same generality as First or Second Laws. This holds for the Laws of Thermodynamics, but also for Newton's Laws.} were known. Hence, for pure atomic or molecular systems, it is meaningful to speak about an absolute entropy. As we shall see below, this is not the case for colloidal systems.

In the final pages of his book on the Principles of Statistical Mechanics, Gibbs discusses the grand-canonical partition function. In this context he comments that there is no need to fix the constant as long as we do not consider exchange of particles between systems. But as soon as we do, the partition function should be divided by $N!$, because otherwise we do not arrive at extensive thermodynamic quantities.  Gibbs writes~\cite{Gibbs}:``{\em ...the principle that the entropy of any body has an arbitrary additive constant is subject to limitation, when different quantities of the same substance are concerned. In this case, the quantity being determined for one quantity of substance, is thereby determined for all quantities of the same substance}''.

Subsequently, with the advent of quantum mechanics, the existence of the factor 1/N! was related to the fact that the square of the quantum-mechanical wave function is invariant  under permutation of identical particles, whereas classically, the permutation of identical particles results in a different configuration in phase space (be it with the same observable properties).   The quantum mechanical indistinguishability of identical particles  is now the standard `explanation' of the Gibbs paradox in most textbooks on Statistical Mechanics. Some, such as Huang~\cite{Huang}, even go as far as stating that the factor N! {\em only} makes sense in the context of quantum mechanics: ``{\em It is not possible to understand classically why we must divide by $N!$ to obtain the correct counting of states. The reason is inherently quantum mechanical}". 

This is not true and several articles have been written that explain that there is no need to invoke quantum mechanics to arrive at the factor $1/N!$. Particularly clear papers have been written on this subject by Jaynes~\cite{Jaynes} and van Kampen~\cite{vanKampen}. The focus of the papers by Jaynes and van Kampen  has been on simple atomic or molecular systems where the concept of identical particles (in the quantum-mechanical sense), still is meaningful. Here I wish to consider the case of systems of particles that, although similar, are {\em all} different. This is the standard situation in colloid science: no two colloids are identical. Even if they would consist of the same number of atoms, their (amorphous) structure differs on a microscopic scale. I should state at the outset that the role of N! in that statistical mechanics of colloidal systems has been discussed by Swendsen~\cite{Swendsen2,Swendsen} and Warren~\cite{Warren}. Much of what I say echoes their comments. 

The key point is that it is perfectly legitimate to apply statistical mechanics to colloidal systems. It is worth analysing why this is the case.  
\section{Mono-disperse colloids}
Let us first consider a somewhat artificial situation where we have a solvent-free colloidal system in zero gravity. Moreover, we assume that the density of the system is so low that we can describe it as an ideal gas. In that case, we can use quantum mechanics to compute the partition function of a system of $N$ `very similar' colloids~\footnote{I use the word `very similar' to indicate that all the observable properties of the colloidal particles -- e.g. mass, size, shape -- are the same, but the microscopic structures of the individual colloids are different.} in a volume $V$ at temperature $T$. We will consider temperatures where the thermal de Broglie wavelength of the colloids is much smaller than the size of the particles. Then we can write the partition function of this system as
\begin{equation}\label{eq:Q0}
Q_{QM}(N,V,T)= (V/\Lambda^3)^N q^N_{int}(T)
\end{equation}
where the subscript `QM' indicates that this is a quantum partition function.  As the colloids are `very similar', I have assumed that they have the same thermal de Broglie wavelength and the same internal partition function $q_{int}(T)$. If we take the logarithm of this purely quantum mechanical partition function, we obtain:
\begin{equation}
\ln Q_{QM}(N,V,T) =  N\ln  (q_{int}(T)/\Lambda^3) + N\ln V
\end{equation}
Clearly, $\ln Q$ is not extensive~\footnote{Free energies thus defined would not be  extensive, but would be additive.} because $\ln Q(2N,2V,T)$ is not equal to $2\ln Q(N,V,T)$ but
\begin{equation}
\ln Q_{QM}(2N,2V,T)= 2 \ln Q_{QM}(N,V,T) + 2N \ln 2 \;.
\end{equation}
Note that quantum indistinguishability cannot fix this non-extensivity because the particles are distinguishable in the quantum sense:  permuting particles is not a symmetry operation on the wave function. 
\section{N! recovered}
Let us next consider two systems, one containing $N_1$ particles in volume $V_1$, the other with $N_2$  particles in volume $V_2$. We prepare the systems  at the same temperature and pressure. This means that (away from a phase transition) the systems have the same density, and hence $N_1/N_2=V_1/V_2$. If we make a small opening in the wall dividing the two systems, mass exchange is possible. We should not expect a net flow of particles between two systems with the same temperature and pressure. Yet, if we write the total partition function of the combined system as
\begin{equation}\label{eq:Product}
Q_{tot}\stackrel{?}{=}Q(N_1,V_1,T)\times Q(N_2,V_2,T)
\end{equation}
then this product is not at a maximum for $N_1/N_2=V_1/V_2$. The underlying problem is that Eqn.~\ref{eq:Product} is wrong. When we consider the total partition function of the combined systems 1 and 2, we must include all possible realisations of the system that result in the same macroscopic state.  That means that we must consider all possible ways in which we can distribute the $N$ colloids, such that there are $N_1$ in volume $V_1$ and $N_2$ in volume $V_2$. That is
\begin{equation}\label{eq:Product2}
Q_{tot}(N_1,V_1,N_2,V_2,T)= {N!\over N_1!N_2!}Q(N_1,V_1,T)\times Q(N_2,V_2,T)\;.
\end{equation}
This point has been made explicitly by Swendsen~\cite{Swendsen2,Swendsen} and Warren~\cite{Warren}. 
If we now differentiate $\ln Q_{tot}$ with respect to $N_1$ at fixed $N$, we get:
\begin{equation}\label{eq:diff1}
\left({\partial \ln Q_{tot}(N_1,V_1,N_2,V_2,T)\over \partial N_1}\right)_N= \left({\partial \ln\left({N!\over N_1!N_2!}Q(N_1,V_1,T)\times Q(N_2,V_2,T)\right)\over \partial N_1}\right)_N\; .
\end{equation}
Using $dN_2=-dN_1$, it is clear that the condition for equilibrium is:
\begin{equation}\label{eq:equi1}
 {\partial \ln\left(Q(N_1,V_1,T)\over N_1!\right)\over \partial N_1}=  {\partial \ln\left(Q(N_2,V_2,T)\over N_2!\right)\over \partial N_2}
 \end{equation}
 Eqn.~\ref{eq:equi1} must express the equality of chemical potential between two phases that are in macroscopically identical states.
 That is:
 \begin{equation}\label{eq:equi2}
 \mu_1=-k_BT{\partial \ln\left(Q(N_1,V_1,T)\over N_1!\right)\over \partial N_1}= -k_BT{\partial \ln\left(Q(N_2,V_2,T)\over N_2!\right)\over \partial N_2}= \mu_2
 \end{equation} 
 In other words, we are forced to conclude that the Helmholtz free energy of a system of `very similar' but distinguishable particles is given by
  \begin{equation}\label{eq:Helmh}
 A(N,V,T)=-k_BT\ln\left(Q(N,V,T)\over N!\right)\;,
 \end{equation} 
although $A(N,V,T)+\alpha N$, with $\alpha$ an arbitrary constant independent of $N$ and $V$ ~\footnote{$\alpha$ can depend linearly on temperature.} would work just as well. 
The factor $N!$ is there, but it is not related to quantum-mechanical indistinguishability but to the fact that permuting very similar colloids does not change the observable properties of the macroscopic systems. What is, and what is not, an observable difference is, in the end, determined by our ability to detect differences. This point was eloquently made by Jaynes~\cite{Jaynes} who discussed the entropy of a mixed system of hypothetical elements (Whifnium, Whafnium and Whoofnium) that are so similar that, initially they were considered as one, then as two and finally as three. The entropy depends on what we know about the system or, more precisely, about what we care to know.   Jaynes writes:  ``{\em ... an illustration of the `anthropomorphic' nature of entropy, would not be apparent to, and perhaps not believed by, someone who thought that entropy  was, like energy, a physical property of the micro state}''. Of course, particles that are  indistinguishable in the quantum sense, meaning  that permuting them is a symmetry operation on the wave function can never be distinguished, no matter what we do, or do not care  to know. 
\begin{figure}[hbt]
  \begin{center}
\includegraphics[width=\textwidth]{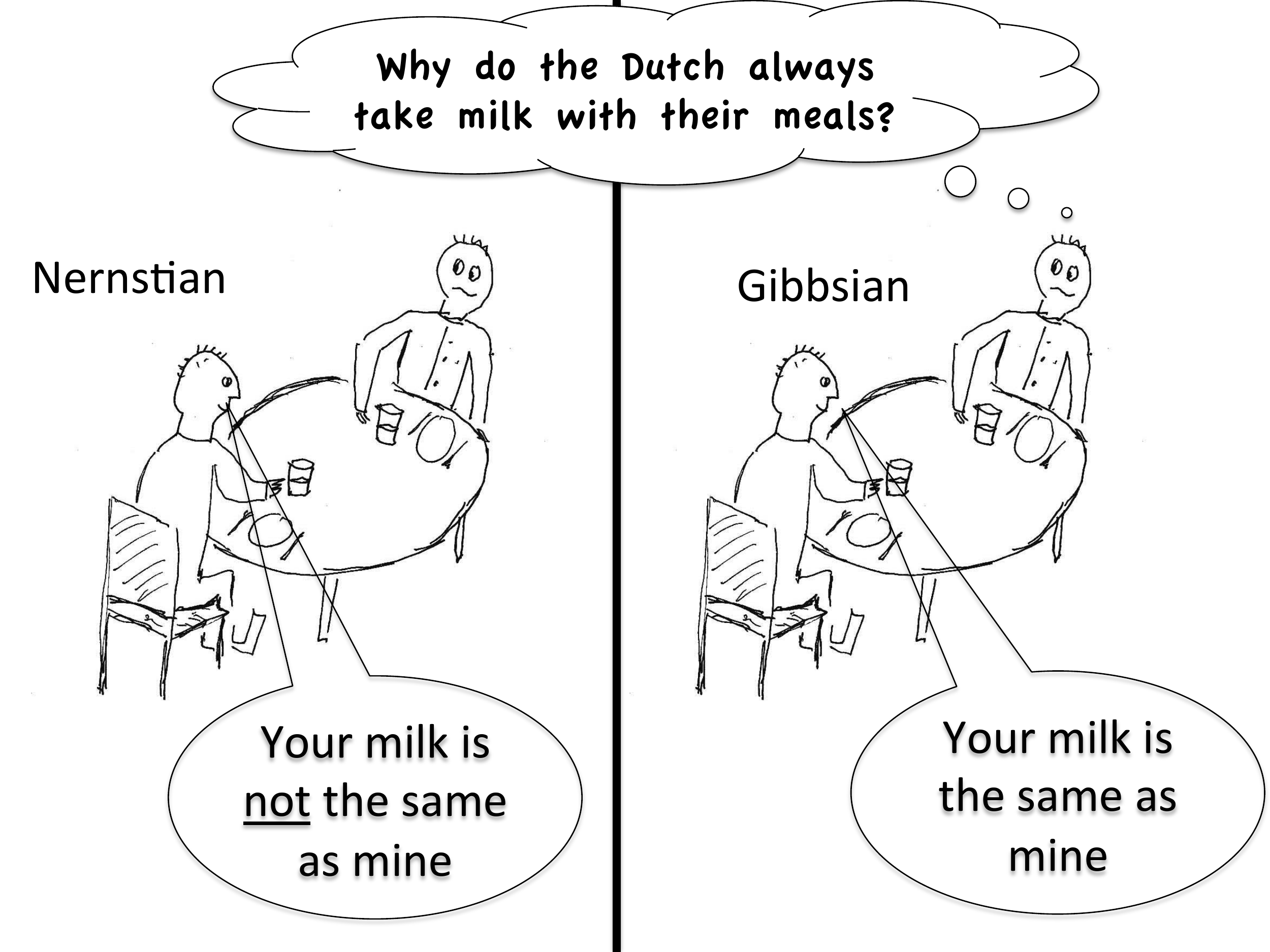}
\caption{The observable properties of two glasses of milk poured from the same bottle are the same. However, all the colloidal particles in the milk are different. Therefore the `quantum' view (left) is that the two systems are {\em not} in the same state, whereas the `thermodynamic' view (right) would be that they are in the same state. Disclaimer: neither Gibbs, nor Nernst ever made the statements above. }%
\label{fig:Milk}
\end{center}
\end{figure}
Yet, if we were to label every individual colloid, then the whole concept of equilibrium under particle exchange becomes meaningless. If we gave every colloid a name, then we would consider the situation where the colloid named `Julius Caesar' is in volume 1  distinct from the situation where this colloid had crossed the Rubicon into volume 2: equilibrium is never possible because every permutation of particles creates a new situation. It is a bit like stamp collectors exchanging stamps that, to the uninitiated, look identical. The experts will consider every distribution of stamps over collectors as a distinct state. 
 
Although I have taken an ideal gas of colloids as an example,  Eqn.~\ref{eq:Helmh} also applies to systems of interacting colloids or, for that matter to any system consisting of large numbers of very similar particles. An immediate consequence of Eqn.~\ref{eq:Helmh} is that we can write for the entropy of our colloidal gas:
 \begin{equation}
S=k_B\ln\left({\Omega(N,V,E)\over N!}\right) \;,
\end{equation}
which is not quite the same as what is written on Boltzmann's grave (but it is perfectly compatible with what Planck wrote in 1901). 

Some readers may feel uncomfortable with the idea of dividing $\Omega$ by $N!$, only to ensure that $\ln (\Omega/N!)$ of a system of very similar particles behaves like the entropy. But that is precisely the procedure that has to be followed in constructing Statistical Mechanics: we start with postulating a correspondence between a computable quantity ($\Omega$) and a thermodynamic quantity that should be at a maximum in a closed system in equilibrium. In the present case, we find that $S=k_B\ln\Omega$ does not do the job, but $S=k_B\ln\left[\Omega(N,V,E)/ N!\right]$ does. 
\section{Polydisperse colloids}
Up to this point, we have assumed that the colloids are so similar that we cannot separate them. However, in practice, colloidal suspensions are usually poly-disperse and we can separate colloids of different sizes by fractionation. The question is how polydispersity will affect the discussion above. As I will show, it makes the absolute value of the entropy meaningless, but the factor $1/N!$ remains. I note that the role of the factor $N!$ in the entropy of polydisperse systems was discussed in a paper by Warren~\cite{Warren} - hence, what I say here is again not very original.

When discussing polydisperse mixtures, the `anthropomorphic' nature of entropy is even more obvious than before. Basically, we have to specify what particles we can separate. The value of the entropy will depend on this choice, but of course, the macroscopic equilibrium will not.

Let us consider an example where we have a colloidal mixture of spherical particles with different sizes. We characterise the size of colloid by its hard-core diameter $\sigma$, and the probability to find a colloid with a size between $\sigma$ and $\sigma+d\sigma$ is given by $P(\sigma)d\sigma$. The first step is to `bin' the particle size distribution into fractions that can be separated. Suppose that there are $m$ such fractions and that the fraction of all particle sizes in bin $i$ that includes all particles with diameters between $\sigma_i$ and $\sigma_{i+1}$  is denoted by $X_i$:
\begin{equation}
X_i=\int_{\sigma_i}^{\sigma_{i+1}} P(\sigma) \; d\sigma
\end{equation}
As we cannot separate particles within one bin, permutations of such particles leave the macroscopic state of the system unaffected. As before, the total number of particles in the combined system (1+2) is denoted by $N$, with $N_1$ particles in system 1 and $N_2$ in system 2.  Then $\omega_{perm}$, the number of ways in which particles can be permuted between systems 1 and 2 is given by
\begin{equation}
\omega_{perm}={\Pi{i=1}^m (N_1(i)+N_2(i))!\over \Pi_i{i=1}^m (N_1(i)!\Pi_{j=1}^mN_2(i)!}
\end{equation}
where $N_{1,2}(i)=N_{1,2}X_i$. 
As before, we require that, in equilibrium,  the total partition function of the combined system must be at a maximum, and that its derivative with respect to all $N_1(i)=N_i-N_2(i)$ must vanish.
The immediate consequence is that the expression for the Helmholtz free energy of a system with $N$ particles in volume $V$ at temperature $T$ must be of the form:
  \begin{equation}\label{eq:HelmhPoly}
 A(N,V,T)=-k_BT\ln\left(Q(N,V,T)\over \Pi_{i=1}^m N(i)!\right)\;,
 \end{equation}
where $N(i) = NX_i$. We can now use the Stirling approximation to write
\begin{equation}
\ln \Pi_{i=1}^m N(i)! =\sum_{i=1}^m   \left(NX_i\ln NX_i -NX_i \right)= \ln N! + N \sum_{i=1}^m X_i\ln X_i\;.
\end{equation}
Let us define an `entropy of mixing' as
\begin{equation}
S_{mix}=-Nk_B\sum_{i=1}^m X_i\ln X_i\;,
\end{equation}
then the expression for the Helmholtz free energy becomes:
  \begin{equation}\label{eq:HelmhPoly2}
 A(N,V,T)=-k_BT\ln\left(Q(N,V,T)\over N!\right) - TS_{mix}\;.
 \end{equation}
Importantly, $S_{mix}$ depends linearly on $N$: it  only changes the reference point for the chemical potential, as long as the composition of the mixture is kept constant.
However, in the case of coexistence between two polydisperse phases with a  different composition, the mixing term does become important. 
It is useful to consider the limit  where the number of bins goes to infinity and the width of the individual bin tends to zero. In that case, we could write~\footnote{Of course, the mathematics here are very sloppy. The way to `read' this equation is to consider that $\int d\sigma (...)$ stands for $\sum \delta\sigma (...)$ with $\delta\sigma$ very small, but sufficiently large that we can apply Stirling's approximation to $\ln (N\delta)!$. After that, we can take the thermodynamic limit...}:
\begin{eqnarray} 
S_{mix}&=&-Nk_B\int_0^\infty d\sigma\; P(\sigma)\ln \left[P(\sigma)\; d\sigma\right]\nonumber\\
 & =& -Nk_B\left[\ln d\sigma+\int_0^\infty d\sigma\; P(\sigma)\ln P(\sigma)\right]\;.
\end{eqnarray}
The second term on the right diverges in the limit $d\sigma\rightarrow 0$. 
However, it is a term that does not depend on composition and is hence immaterial for phase coexistence. We can ignore it.
The physically meaningful part of $S_{mix}$ is:
\begin{equation} \label{eq:Spoly}
S_{mix}=-Nk_B\int_0^\infty d\sigma\; P(\sigma)\ln P(\sigma)\;,
\end{equation} 
which is well defined~\footnote{Almost: if we change the integration variable from $\sigma$ to a monotonic function of $\sigma$, a Jacobian will enter into Eqn.\ref{eq:Spoly}, but not in any of the entropy differences that are important for phase equilibria.}.
\section{Conclusions}
Of course, the expression $S=k_B\ln\Omega$ {\em is} valid for a system consisting of indistinguishable quantum particles: $\Omega(E,V,N)$ counts the number of eigenstates with energy $E$ and permutations of particles will map a given  eigenstate onto itself (possibly with a minus sign). However, as soon as we deal with distinguishable colloids, we need to divide the partition function by $N!$ in order to obtain an extensive Helmholtz free energy. In that case, $S= k_B\ln(\Omega/N!) = k_B\ln\Omega +\mbox{const.}$, the very expression that Planck wrote down in 1901.
\section*{Acknowledgments}
This work has been supported by the
EPSRC grant $N^{\circ}$ EP/I000844/1 and ERC Advanced Grant 227758.
I gratefully acknowledge heated discussions with Fabien Paillusson and Daniel Asenjo. 
I thank Patrick Warren for sending me Ref.~\cite{Warren}. 
I thank Erika Eiser for carefully reading the manuscript.

\end{document}